\documentclass[preprint,showpacs,showkeys,preprintnumbers,amsmath,amssymb]{revtex4}
\usepackage{graphics}
\begin{document}
\title{Attenuation and damping of electromagnetic fields:\\Influence of inertia and displacement current}
\author{F. E. M. Silveira}
\email{francisco.silveira@ufabc.edu.br}
\affiliation{Centro de Ci\^encias Naturais e Humanas, Universidade Federal do ABC, Rua Santa Ad\'elia, 166, Bairro Bangu, CEP 09210-170, Santo Andr\'e, SP, Brazil}
\author{J. A. S. Lima}
\email{limajas@astro.iag.usp.br}
\affiliation{Instituto de Astronomia, Geof\'\i sica e Ci\^encias Atmosf\'ericas, Universidade de S\~ao Paulo, Rua do Mat\~ao, 1226, Cidade Universit\'aria, CEP 05508-900, S\~ao Paulo, SP, Brazil}

\begin{abstract}
New results for attenuation and damping of electromagnetic fields in rigid conducting media are derived under the conjugate influence of inertia due to charge carriers and displacement current. Inertial effects are described by a relaxation time for the current density in the realm of an extended Ohm's law. The classical notions of poor and good conductors are rediscussed on the basis of an effective electric conductivity, depending on both wave frequency and relaxation time. It is found that the attenuation for good conductors at high frequencies depends solely on the relaxation time. This means that the penetration depth saturates to a minimum value at sufficiently high frequencies. It is also shown that the actions of inertia and displacement current on damping of magnetic fields are opposite to each other. That could explain why the classical decay time of magnetic fields scales approximately as the diffusion time. At very small length scales, the decay time could be given either by the relaxation time or by a fraction of the diffusion time, depending whether inertia or displacement current, respectively, would prevail on magnetic diffusion.
\end{abstract}

\pacs{41.20.-q; 41.20.Gz; 41.20.Jb}
\keywords{boundary value problems; electromagnetic shielding; quasi static electromagnetic fields}
\maketitle

\section{introduction}
It is widely accepted that, under ordinary circumstances, the standard Ohm's law,
\begin{eqnarray}
\vec{J}=\sigma\vec{E},
\label{1}
\end{eqnarray}
which relates the electric field $\vec{E}$ to the current density $\vec{J}$ flowing in a linear, isotropic and homogenous medium with electric conductivity $\sigma$, provides a very good description of electrodynamic processes occurring in conductors. However, for gases and plasmas, or even for high frequency electromagnetic phenomena, the possibility of non Ohmic features has attracted the interest of many researchers \cite{Wannier,Llebot,Krall,Boyd,Jou88,Jou96}. In those situations, a generalized expression for Ohm's law has been proposed,
\begin{eqnarray}
\left(1+\tau\frac{d}{dt}\right)\vec{J}=\sigma\vec{E},
\label{2}
\end{eqnarray}
where $\tau$ is a constant with dimension of time and the assigned derivative is the total time derivative. The above generalized law can be deduced rigorously, either on the basis of kinetic theory \cite{Krall,Boyd} or in the framework of extended irreversible thermodynamics \cite{Jou88,Jou96,Israel}. Clearly, in the limit $\tau\rightarrow0$, the standard Ohm's law is recovered. Still more interestingly, the current density does not immediately vanish as soon as the electric field is turned off. As a consequence of inertial effects due to charge carriers in the conductor, it approaches zero in a time scale which is determined by $\tau$, thereby quantifying the relaxation time of the current density (see Appendix A).

In principle, the above time dependent correction added to Ohm's law is negligible under stationary conditions, as well as for low frequency regimes. However, it may have several interesting physical consequences to the behavior of transient electromagnetic fields, and also at high frequency limits. In particular, as discussed in \cite{Cuevas}, it should modify considerably the classical results for space attenuation of electromagnetic waves \cite{Jackson} and time damping of magnetic fields \cite{Landau} in rigid conducting media.

In this work, new results for attenuation and damping of electromagnetic fields in rigid conducting media are derived under the combined action of inertia and displacement current. The classical conceptions of poor and good conductors are extended with basis on an effective electric conductivity, which depends on both wave frequency and relaxation time. It is shown that the attenuation for good conductors at high frequency electromagnetic waves depends on the relaxation time only, that is, the penetration depth must saturate to a minimum value at sufficiently high frequencies. Such a result leads to the possibility of measurement of the relaxation time. It is also found that the influences of displacement current and inertia on damping of magnetic fields are opposed to each other. That could explain why, under normal conditions, the decay time of magnetic fields is given approximately by the diffusion time. At very small length scales (at the nano scale, for instance), the decay time could be given either by a fraction of the diffusion time or by the relaxation time, depending whether displacement current or inertia, respectively, would prevail on magnetic diffusion. However, it should be hardly observable in the latter case, under ordinary circumstances, given the smallness of the length scale.

The paper is organized as follows. In Sec. II, we cast the basic equations for describing the influence of inertia and displacement current on attenuation and damping. In Sec. III, the attenuation of electromagnetic waves is discussed. The above mentioned effective conductivity is introduced, and an expression for the ratio of time averaged magnetic to electric energy densities is derived in terms of both wave frequency and relaxation time. Then, the low and high frequency regimes are explored for poor and good conductor limits.  In Sec. IV, the damping of magnetic fields is discussed. In particular, it is shown that the actions of inertia and displacement current on damping are opposite to each other, and the decay time of magnetic fields is explored at very small length scales. In Sec. V, the main conclusions are summarized. 

\section{basic equations}
For a rigid conductor, except for the small massive charge carriers, there is no relative motion between its constitutive parts. In that situation, the total time derivative in Eq. (\ref{2}) becomes a partial time derivative and the generalized Ohm's law takes the form
\begin{eqnarray}
\left(1+\tau\frac{\partial}{\partial t}\right)\vec{J}=\sigma\vec{E}.
\label{3}
\end{eqnarray}
By combining Eq. (\ref{3}) with Maxwell's equations (recall there is no free charge in the bulk of the conducting medium),
\begin{eqnarray}
\nabla\times\vec{B}&=&\mu\epsilon\frac{\partial\vec{E}}{\partial t}+\mu\vec{J},\;\;\;\;\;\;\;\;\;\;\nabla\cdot\vec{B}=0,\nonumber\\\nabla\times\vec{E}&=&-\frac{\partial\vec{B}}{\partial t},\;\;\;\;\;\;\;\;\;\;\;\;\;\;\;\;\;\;\;\;\nabla\cdot\vec{E}=0,
\label{4}
\end{eqnarray}
where $\mu$ and $\epsilon$ are the magnetic permeability and electric permittivity, respectively, of the medium, one may easily check that the space and time evolution of the magnetic field $\vec{B}$ is governed by
\begin{eqnarray}
\left(1+\tau\frac{\partial}{\partial t}\right)\nabla^2\vec{B}=\left(1+\tau\frac{\partial}{\partial t}\right)\mu\epsilon\frac{\partial^2\vec{B}}{\partial t^2}+\mu\sigma\frac{\partial\vec{B}}{\partial t}.
\label{5}
\end{eqnarray}
In the limit $\tau\rightarrow0$, the classical space and time evolution of the magnetic field in a rigid conductor is recovered \cite{Jackson}.

If we assume that both magnetic and electric fields, as well as the current density, vary in space and time as $\sim\exp{\left(\imath\vec{k}\cdot\vec{x}-\imath\omega t\right)}$, where $\vec{k}$ and $\omega$ are the complex propagation vector and real angular frequency, respectively, of a monochromatic electromagnetic wave, Eqs. (\ref{3}) and (\ref{5}) lead to
\begin{eqnarray}
\left(1-\imath\omega\tau\right)\vec{J}_0=\sigma\vec{E}_0,
\label{6}
\end{eqnarray}
\begin{eqnarray}
\left(1-\imath\omega\tau\right)k^2=\left(1-\imath\omega\tau\right)\omega^2\mu\epsilon+\imath\omega\mu\sigma,
\label{7}
\end{eqnarray}
respectively, where $\vec{J}_0$ and $\vec{E}_0$ are the complex vector amplitudes of the current density and electric field, respectively, and
$k=\left(\vec{k}\cdot\vec{k}\right)^{1/2}$ is the complex mode number of the electromagnetic wave. In the limit $\tau\rightarrow0$, the corresponding classical results in Fourier space are recovered \cite{Jackson}. Eqs. (\ref{3}) to (\ref{7}) are the basic equations for describing the influence of inertia and displacement current on space attenuation of electromagnetic waves and time damping of magnetic fields in rigid conducting media.

\section{space attenuation of electromagnetic waves}

\subsection{General considerations}

\subsubsection{Effective conductivity and attenuation factor}
As it appears, Eq. (\ref{6}) gives a linear algebraic relation between the complex vector amplitudes $\vec{J}_0$ and $\vec{E}_0$. For our purposes, we rewrite it in a more convenient manner,
\begin{eqnarray}
\vec{J}_0=\sigma_{\rm eff}\exp{\left(\imath\varphi\right)}\vec{E}_0,
\label{8}
\end{eqnarray}
where we introduce the effective electric conductivity of the rigid medium,
\begin{eqnarray}
\sigma_{\rm eff}=\frac{\sigma}{\sqrt{1+\omega^2\tau^2}},
\label{9}
\end{eqnarray}
and the time dephasing angle of the current density $\vec{J}$ with respect to the electric field $\vec{E}$,
\begin{eqnarray}
\varphi=\tan^{-1}\left(\omega\tau\right).
\label{10}
\end{eqnarray}
The usual definitions of poor and good conductors \cite{Jackson} correspond to the limits $\sigma\ll\epsilon\omega$ and $\sigma\gg\epsilon\omega$, respectively. With basis on Eqs. (\ref{8}) to (\ref{10}), we see that, when inertial effects are taken into account, those limits must be replaced by
$\sigma_{\rm eff}\ll\epsilon\omega$ and $\sigma_{\rm eff}\gg\epsilon\omega$, respectively. Indeed, in the limit $\tau\rightarrow0$, the effective electric conductivity takes its ordinary value, and the current density and electric field have the same time phase angle. As a consequence, the extended limits recover the corresponding classical ones. Moreover, we note that those extensions are in accordance with the observation that the conductivity of the medium generally depends on the frequency of the wave \cite{Jackson}. { However, in still more general situations, like those involving anisotropic media, it is well established that the electric conductivity must be treated as a tensor \cite{Bladel}. In these cases, Eq. (\ref{2}) might be extended to
\begin{eqnarray}
\left(1+\tau\frac{d}{dt}\right)\vec{J}=\stackrel{\leftrightarrow}\sigma\vec{E},
\label{11}
\end{eqnarray}
where $\stackrel{\leftrightarrow}\sigma$ denotes the electric conductivity tensor. With basis on Eq. (\ref{11}), we see that all classical frequency regimes and conduction limits should require revisions due to possible introductions of other physically relevant frequency and time scales.}

Besides, Eq. (\ref{7}) yields the dispersion relation between the complex mode number and real angular frequency of the electromagnetic wave,
\begin{eqnarray}
k^2=\frac{\omega^2\mu\epsilon}{1+\omega^2\tau^2}\left(1-\frac{\sigma\tau}{\epsilon}+\omega^2\tau^2+\imath\frac{\sigma}{\epsilon\omega}\right).
\label{12}
\end{eqnarray}
Following standard lines \cite{Jackson}, we write the complex mode number as $k=\beta+\imath\alpha/2$, so that its real and imaginary parts are given by
\begin{eqnarray}
\left.
\begin{array}{c}
\beta \\ \alpha/2
\end{array}
\right\}=\omega\sqrt{\frac{\mu\epsilon/2}{1+\omega^2\tau^2}}\left[\sqrt{\left(1-\frac{\sigma\tau}{\epsilon}+\omega^2\tau^2\right)^2+\left(\frac{\sigma}{\epsilon\omega}\right)^2}\pm\left(1-\frac{\sigma\tau}{\epsilon}+\omega^2\tau^2\right)\right]^{1/2}.
\label{13}
\end{eqnarray}
The imaginary part of $k$, $\alpha/2$, is known as the attenuation factor of the electromagnetic wave propagating in the rigid conductor \cite {Jackson}. Eqs. (\ref{13}) are the same as Eqs. (11) of \cite{Cuevas}. Those authors have studied the problem of attenuation of electromagnetic waves in rigid conductors for the low frequency regime, $\omega\tau\ll1$. In this article, we rediscuss their results but also explore Eqs. (\ref{13}) in the high frequency regime, $\omega\tau\gg1$, on the basis of the effective electric conductivity, introduced above. Moreover, we include an analysis of the time averaged electromagnetic energy density in the discussion, for all different frequency regimes and conduction limits.

\subsubsection{Transverse wave and energy density}
To begin with, we observe that the space and time fluctuation of the electromagnetic field, assigned above, together with Maxwell's equations, Eqs. (\ref{4}), lead to
\begin{eqnarray}
k\hat{n}\times\vec{E}_0=\omega\vec{B}_0,\;\;\;\;\;\;\;\;\;\;\hat{n}\cdot\vec{E}_0=0,
\label{14}
\end{eqnarray}
where $\hat{n}$ is the real unit vector in the direction of the complex propagation vector $\vec{k}$, and $\vec{B}_0$ is the complex vector amplitude of the magnetic field, thereby showing the expected transverse nature of the electromagnetic wave. By writing the complex mode number as
$k=\mid k\mid\exp\left(\imath\phi\right)$, it follows that the ratio of the real scalar amplitude of $\vec{B}$ to that of $\vec{E}$ is given by
\begin{eqnarray}
\frac{\mid\vec{B}_0\mid}{\mid\vec{E}_0\mid}=\frac{\mid k\mid}{\omega},
\label{15}
\end{eqnarray}
and $\vec{B}$ lags $\vec{E}$ in time by the phase angle $\phi$. From Eqs. (\ref{13}), one may easily check that
\begin{eqnarray}
\mid k\mid=\omega\sqrt{\frac{\mu\epsilon}{1+\omega^2\tau^2}}\left[\left(1-\frac{\sigma\tau}{\epsilon}+\omega^2\tau^2\right)^2+\left(\frac{\sigma}{\epsilon\omega}\right)^2\right]^{1/4},
\label{16}
\end{eqnarray}
\begin{eqnarray}
\phi=\tan^{-1}\frac{\epsilon\omega}{\sigma}\left[\sqrt{\left(1-\frac{\sigma\tau}{\epsilon}+\omega^2\tau^2\right)^2+\left(\frac{\sigma}{\epsilon\omega}\right)^2}-\left(1-\frac{\sigma\tau}{\epsilon}+\omega^2\tau^2\right)\right].
\label{17}
\end{eqnarray}
In the limit $\tau\rightarrow0$, Eqs. (\ref{16}) and (\ref{17}) recover the classical results for $\mid k\mid$ and $\phi$ \cite{Jackson}.

In this work, we take a step further by considering the time average of magnetic and electric energies per unit volume of the medium,
\begin{eqnarray}
\bar{u}_B=\frac{\mid\vec{B}_0\mid^2}{2\mu},\;\;\;\;\;\;\;\;\;\;\bar{u}_E=\frac{\epsilon\mid\vec{E}_0\mid^2}{2},
\label{18}
\end{eqnarray}
respectively. First, from Eq. (\ref{15}), we see that the ratio of time averaged magnetic to electric energy densities is given by
\begin{eqnarray}
\frac{\bar{u}_B}{\bar{u}_E}=\frac{\mid k\mid^2}{\omega^2\mu\epsilon}.
\label{19}
\end{eqnarray}
Then, from Eq. (\ref{16}), it follows that Eq. (\ref{19}) can be written as
\begin{eqnarray}
\frac{\bar{u}_B}{\bar{u}_E}=\frac{1}{1+\omega^2\tau^2}\sqrt{\left(1-\frac{\sigma\tau}{\epsilon}+\omega^2\tau^2\right)^2+\left(\frac{\sigma}{\epsilon\omega}\right)^2}.
\label{20}
\end{eqnarray}
In the limit $\tau\rightarrow0$, one may easily check that the expected classical result for the ratio $\bar{u}_B/\bar{u}_E$ is recovered \cite{Jackson}. However, in the high frequency regime, $\omega\tau\gg1$, a crude inspection of Eq. (\ref{20}) suggests that $\bar{u}_B$ always approaches $\bar{u}_E$. Somewhat surprisingly, that ceases to be true in the good conductor limit, as will be shown on the basis of the effective electric conductivity, introduced above.

\subsection{Low frequency regime}
The low frequency regime, $\omega\tau\ll1$, may be attained by considering terms of $\mathcal{O}\left(\omega^2\tau^2\right)=0$. In this order of approximation, we see at once that $\sigma_{\rm eff}=\sigma$, although $\vec{J}$ is slightly dephased in time with respect to $\vec{E}$ by the small angle $\varphi=\omega\tau$. The real and imaginary parts of the complex mode number are
\begin{eqnarray}
\left.
\begin{array}{c}
\beta \\ \alpha/2
\end{array}
\right\}=\omega\sqrt{\frac{\mu\epsilon}{2}}\left[\sqrt{\left(1-\frac{\sigma\tau}{\epsilon}\right)^2+\left(\frac{\sigma}{\epsilon\omega}\right)^2}\pm\left(1-\frac{\sigma\tau}{\epsilon}\right)\right]^{1/2}.
\label{21}
\end{eqnarray}
Eqs. (\ref{21}) should be compared with Eqs. (12) of \cite{Cuevas}. As it appears, those authors have neglected the term of $\mathcal{O}\left(\tau^2\right)$ under the square root sign. However, this term could not have been neglected since the square root implies it is of $\mathcal{O}\left(\tau\right)$ effectively. Indeed, it will be shown that by consistently retaining it both classical results in the poor and good conductor limits are corrected to terms of $\mathcal{O}\left(\omega\tau\right)$.

The modulus of the complex mode number is
\begin{eqnarray}
\mid k\mid=\omega\sqrt{\mu\epsilon}\left[\left(1-\frac{\sigma\tau}{\epsilon}\right)^2+\left(\frac{\sigma}{\epsilon\omega}\right)^2\right]^{1/4},
\label{22}
\end{eqnarray}
the magnetic field lags the electric field in time by the phase angle
\begin{eqnarray}
\phi=\tan^{-1}\frac{\epsilon\omega}{\sigma}\left[\sqrt{\left(1-\frac{\sigma\tau}{\epsilon}\right)^2+\left(\frac{\sigma}{\epsilon\omega}\right)^2}-\left(1-\frac{\sigma\tau}{\epsilon}\right)\right],
\label{23}
\end{eqnarray}
and the ratio of time averaged magnetic to electric energy densities is
\begin{eqnarray}
\frac{\bar{u}_B}{\bar{u}_E}=\sqrt{\left(1-\frac{\sigma\tau}{\epsilon}\right)^2+\left(\frac{\sigma}{\epsilon\omega}\right)^2}.
\label{24}
\end{eqnarray}
Next, we discuss the limits of Eqs. (\ref{21}) to (\ref{24}) for poor and good conductors.

\subsubsection{Poor conductor limit}
Poor conductors at low frequency electromagnetic waves are described by requiring the condition $\sigma\ll\epsilon\omega$ to be satisfied. In this approximation,
\begin{eqnarray}
\mid k\mid\cong\beta=\omega\sqrt{\mu\epsilon}-\frac{\sigma}{2}\sqrt{\frac{\mu}{\epsilon}}\omega\tau,\;\;\;\;\;\;\;\;\;\;\frac{\alpha}{2}=\frac{\sigma}{2}\sqrt{\frac{\mu}{\epsilon}}.
\label{25}
\end{eqnarray}
We see that, although the small classical value for the attenuation factor is not altered by inertial effects due to charge carriers, the real part of the complex mode number, as well as its own modulus, slightly decrease as a consequence of consistently retaining the terms of $\mathcal{O}\left(\tau\right)$ in Eqs. (\ref{21}). The first of Eqs. (\ref{25}) corrects the result of \cite{Cuevas}. As already mentioned, those authors have neglected the term of $\mathcal{O}\left(\tau^2\right)$ under the square root sign in their Eqs. (12). That is the origin of their result.

The magnetic field lags the electric field in time by the same small classical phase angle,
\begin{eqnarray}
\phi=\frac{\sigma}{2\epsilon\omega}.
\label{26}
\end{eqnarray}
Interestingly, the time averaged magnetic and electric energy densities are no longer identical, as they were classically, since the ratio $\bar{u}_B/\bar{u}_E$ slightly decreases with respect to unit,
\begin{eqnarray}
\frac{\bar{u}_B}{\bar{u}_E}=1-\frac{\sigma\tau}{\epsilon}.
\label{27}
\end{eqnarray}
In the limit $\tau\rightarrow0$, the first of Eqs. (\ref{25}), and (\ref{27}) recover the expected classical results \cite{Jackson}.

\subsubsection{Good conductor limit}
Good conductors at low frequency electromagnetic waves are described by requiring the condition $\sigma\gg\epsilon\omega$ to be satisfied. In this approximation,
\begin{eqnarray}
\left.
\begin{array}{c}
\beta \\ \alpha/2
\end{array}
\right\}=\sqrt{\frac{\mu\sigma\omega}{2}}\left(1\mp\frac{\omega\tau}{2}\right).
\label{28}
\end{eqnarray}
We see that the classical values of $\beta$ and $\alpha/2$ are decreased and increased, respectively, by the same small number, $\omega\tau/2$. Eqs. (\ref{28}) are essentially the same as Eqs. (13) of \cite{Cuevas}, provided the binomial expansions of the expressions $\sqrt{1\mp\omega\tau}$ are carried on to the term of $\mathcal{O}\left(\omega\tau\right)\ll1$. The classical result for the modulus of the complex mode number, $\mid k\mid=\sqrt{\mu\sigma\omega}$, is not altered by inertial effects.

Interestingly, the classical $\pi/4$ time dephasing angle of $\vec{B}$ with respect to $\vec{E}$ is slightly increased by inertial effects, since
\begin{eqnarray}
\phi=\frac{\pi}{4}+\frac{\omega\tau}{2}.
\label{29}
\end{eqnarray}
As in the classical case, the time averaged electromagnetic energy density is almost entirely magnetic in nature, since the ratio $\bar{u}_B/\bar{u}_E$ is given by the large number
\begin{eqnarray}
\frac{\bar{u}_B}{\bar{u}_E}=\frac{\sigma}{\epsilon\omega}.
\label{30}
\end{eqnarray}
In the limit $\tau\rightarrow0$, Eqs. (\ref{28}) and (\ref{29}) recover the expected classical results \cite{Jackson}.

\subsection{High frequency regime}
Now, we explore a situation which, to the best of our knowledge, had not been considered elsewhere, namely, the influence of inertia on attenuation at high frequencies. The high frequency regime may be achieved by requiring the condition $\omega\tau\gg1$ to be satisfied. In this approximation, we see at once that the classical electric conductivity is strongly suppressed by inertial effects, given that
\begin{eqnarray}
\sigma_{\rm eff}=\frac{\sigma}{\omega\tau},
\label{31}
\end{eqnarray}
whilst $\vec{J}$ and $\vec{E}$ are completely out of phase in time, since
\begin{eqnarray}
\varphi=\frac{\pi}{2}.
\label{32}
\end{eqnarray}
In order to simplify the checking by the reader of the limits for poor and good conductors, the following formulae are written in terms of $\sigma_{\rm eff}$, instead of $\sigma$.

The real and imaginary parts of the complex mode number are
\begin{eqnarray}
\left.
\begin{array}{c}
\beta \\ \alpha/2
\end{array}
\right\}=\sqrt{\frac{\mu\epsilon\omega}{2\tau}}\left[\sqrt{\left(\frac{\sigma_{\rm eff}\tau}{\epsilon}-\omega\tau\right)^2+\left(\frac{\sigma_{\rm eff}}{\epsilon\omega}\right)^2}\mp\left(\frac{\sigma_{\rm eff}\tau}{\epsilon}-\omega\tau\right)\right]^{1/2},
\label{33}
\end{eqnarray}
while its modulus is given by
\begin{eqnarray}
\mid k\mid=\sqrt{\frac{\mu\epsilon\omega}{\tau}}\left[\left(\frac{\sigma_{\rm eff}\tau}{\epsilon}-\omega\tau\right)^2+\left(\frac{\sigma_{\rm eff}}{\epsilon\omega}\right)^2\right]^{1/4}.
\label{34}
\end{eqnarray}
The magnetic field lags the electric field in time by the phase angle
\begin{eqnarray}
\phi=\tan^{-1}\frac{\epsilon\omega}{\sigma_{\rm eff}}\left[\sqrt{\left(\frac{\sigma_{\rm eff}\tau}{\epsilon}-\omega\tau\right)^2+\left(\frac{\sigma_{\rm eff}}{\epsilon\omega}\right)^2}+\left(\frac{\sigma_{\rm eff}\tau}{\epsilon}-\omega\tau\right)\right],
\label{35}
\end{eqnarray}
and the ratio of time averaged magnetic to electric energy densities satisfies
\begin{eqnarray}
\frac{\bar{u}_B}{\bar{u}_E}=\frac{1}{\omega\tau}\sqrt{\left(\frac{\sigma_{\rm eff}\tau}{\epsilon}-\omega\tau\right)^2+\left(\frac{\sigma_{\rm eff}}{\epsilon\omega}\right)^2}.
\label{36}
\end{eqnarray}
Next, we discuss the limits of Eqs. (\ref{33}) to (\ref{36}) for poor and good conductors.

\subsubsection{Poor conductor limit}
Poor conductors at high frequency electromagnetic waves are described by requiring the condition $\sigma_{\rm eff}\ll\epsilon\omega$ to be satisfied. In this approximation,
\begin{eqnarray}
\beta=\omega\sqrt{\mu\epsilon},\;\;\;\;\;\;\;\;\;\;\frac{\alpha}{2}=\frac{1}{\omega^2\tau^2}\frac{\sigma}{2}\sqrt{\frac{\mu}{\epsilon}}.
\label{37}
\end{eqnarray}
We see that the already small attenuation factor for poor conductors at low frequencies is strongly suppressed by inertial effects at high frequencies. The modulus of the complex mode number for poor conductors at low frequencies is also suppressed by inertial effects at high frequencies,
\begin{eqnarray}
\mid k\mid=\frac{\omega\sqrt{\mu\epsilon}}{\sqrt{\omega\tau}},
\label{38}
\end{eqnarray}
although less strongly than the attenuation factor.

The already small time dephasing angle of $\vec{B}$ with respect to $\vec{E}$ for poor conductors at low frequencies is again strongly suppressed by inertial effects at high frequencies, given that
\begin{eqnarray}
\phi=\frac{1}{\omega^2\tau^2}\frac{\sigma}{2\epsilon\omega}.
\label{39}
\end{eqnarray}
The time averaged magnetic and electric energy densities are almost identical for poor conductors at high frequencies, since $\bar{u}_B\cong\bar{u}_E$. That is the same result which holds for poor conductors at low frequencies \cite{Jackson}.

\subsubsection{Good conductor limit}
Good conductors at high frequency electromagnetic waves are described by requiring the condition $\sigma_{\rm eff}\gg\epsilon\omega$ to be satisfied. In this approximation,
\begin{eqnarray}
\beta=\frac{\sqrt{\mu\sigma/\tau}}{2\omega\tau},\;\;\;\;\;\;\;\;\;\;\mid k\mid\cong\frac{\alpha}{2}=\sqrt{\frac{\mu\sigma}{\tau}}.
\label{40}
\end{eqnarray}
Interestingly, we see that the large attenuation factor depends on the relaxation time only. { The penetration depth of the electromagnetic wave within the conductor, but close to its external surface (skin effect), is defined as $\delta=2/\alpha$ \cite{Jackson}. By neglecting inertial effects, the classical value of $\delta$ for a good conductor scales as $\sim1/\sqrt{\omega}$ [see the second of Eqs. (\ref{28})]. In other words, it should diminish without limit at high frequency electromagnetic waves. However, our new result, the second of Eqs. (\ref{40}), ensures that $\delta$ does not depend on the frequency of the electromagnetic wave, provided inertial effects are taken into account. The physical meaning of this is that $\delta$ must saturate to a minimum value at sufficiently high frequencies. By making the substitutions $\tau\rightarrow\tau_{\rm c}$ and $\sigma\rightarrow\sigma_{\rm D}$ (see Appendix A), we see that, for a typical good metal like copper ($\tau_{\rm c}\cong2.7\times10^{-14}\;{\rm s}$), $\delta\cong1.9\times10^{-6}\;{\rm cm}$. At this point, we would like to encourage the experimentalists to try and find the inertial relaxation time of the current density by measuring the penetration depth for a good conductor at high frequencies, which is within the limits of present day laboratory capabilities (see Appendix B).}

The magnetic and electric fields are almost completely out of phase in time, since $\phi\cong\pi/2$. Once $\varphi\cong\pi/2$ too, the time dephasing angle of $\vec{J}$ with respect to $\vec{B}$ is approximately an integral multiple of $\pi$. In other words, $\vec{J}$ and $\vec{B}$ are almost either at the same or at opposite time phase angles. As already mentioned, an interesting consequence of the introduction of the notion of effective electric conductivity in terms of wave frequency and relaxation time is that the time averaged electromagnetic energy density is almost entirely magnetic in nature for good conductors at high frequencies, as long as the ratio $\bar{u}_B/\bar{u}_E$ is given by the large number
\begin{eqnarray}
\frac{\bar{u}_B}{\bar{u}_E}=\frac{1}{\omega\tau}\frac{\sigma}{\epsilon\omega}.
\label{41}
\end{eqnarray}
Although numerically different, that is the same qualitative result which holds for good conductors at low frequencies \cite{Jackson}.

\begin{figure}
\resizebox{4in}{4in}{\includegraphics{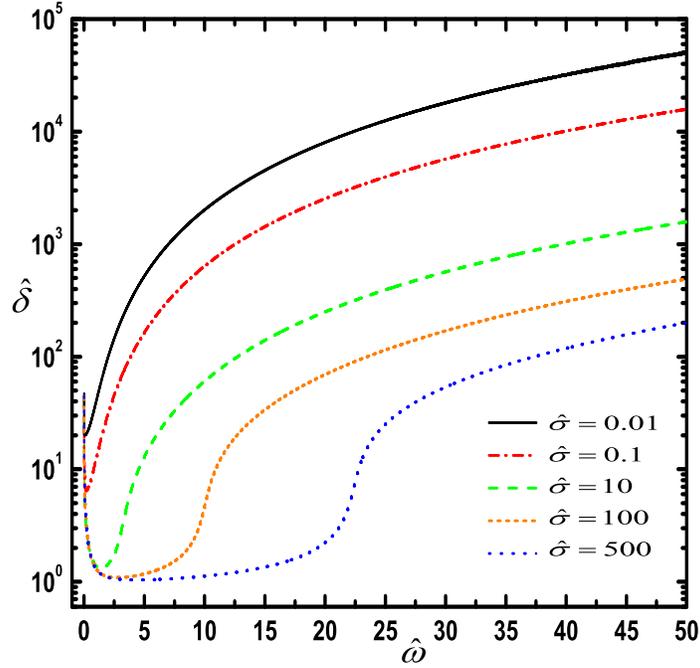}}
\caption{\label{fig_1} Normalized penetration depth as a function of $\hat{\omega}$ and some selected values of $\hat{\sigma}$. Note the tendency for saturation of $\hat{\delta}$ when $\hat{\sigma}$ increases.}
\end{figure}

\subsection{General case: penetration depth}
{In order to discuss the general case, let us introduce dimensionless quantities describing the attenuation in conducting media. As one may check, by defining  
\begin{eqnarray}
\hat{\sigma}=\frac{\sigma}{\epsilon/\tau},\;\;\;\;\;\;\;\;\;\;\hat{\omega}=\frac{\omega}{1/\tau},\;\;\;\;\;\;\;\;\;\;\hat{\delta}=\frac{\delta}{\sqrt{\tau/\left(\mu\sigma\right)}},
\label{42}
\end{eqnarray}
the second of Eqs. (\ref{13}) can be rewritten as 
\begin{eqnarray}
\hat{\delta}=\sqrt{\frac{2\hat{\sigma}\left(1+\hat{\omega}^2\right)}{\hat{\omega}}}\left[\sqrt{\left(1-\hat{\sigma}+\hat{\omega}^2\right)^2\hat{\omega}^2+\hat{\sigma}^2}-\left(1-\hat{\sigma}+\hat{\omega}^2\right)\hat{\omega}\right]^{-1/2}.
\label{43}
\end{eqnarray}

In Figure 1 we show the behavior of the normalized penetration depth as a function of $\hat{\omega}$ and some selected values of $\hat{\sigma}$.
As remarked earlier, in the high frequency limit the penetration depth saturates in the case of good conductors.}

\section{time damping of magnetic fields}
In the second part of this paper, we explore a problem which is closely related to the description of space attenuation of electromagnetic waves, namely, the estimate of time damping of magnetic fields in rigid conducting media. The classical approach to the latter starts from Eq. (\ref{5}) in the limit $\tau\rightarrow0$, and in absence of displacement current. In other words, inertial effects are neglected, at low frequencies. The resulting equation is a diffusion type equation for the magnetic field, where $\left(\mu\sigma\right)^{-1}$ plays the role of a diffusion coefficient. The physical situation one wants to describe is that of a rigid conducting medium subjected to an external magnetic field, which is suddenly removed. The time dependence of each eigenfunction of the diffusion equation is ascribed as $\sim\exp\left(-\gamma_lt\right)$, where the associated eigenvalue $\gamma_l$ is interpreted as a damping factor. The decay time of the damping field is estimated as $T\sim1/\gamma_1$, where $\gamma_1={\rm min}\left\{\gamma_l\right\}$, by simultaneously estimating $\mid\nabla^2\mid\sim1/L^2$, with $L$ denoting a characteristic length scale of the sample. The classical result for the decay time is given by the diffusion time, $T=\mu\sigma L^2$. This approach is commonly referred to as the quasi static approximation \cite{Landau}.

In this article, we follow an alternative path to tackle the above mentioned problem. The idea is that, since the space and time dependence of the external magnetic field may be always Fourier analyzed, we can start at once from Eq. (\ref{7}) by making the substitution $\omega\rightarrow-\imath\gamma$, where $\gamma$ is interpreted as a real damping factor associated to the real mode number $k$ of a given Fourier component. As a consequence, the time dependence of each component takes the desired form, $\sim\exp\left(-\gamma t\right)$, and we get
\begin{eqnarray}
\left(1-\gamma\tau\right)k^2=\gamma\mu\sigma-\left(1-\gamma\tau\right)\gamma^2\mu\epsilon.
\label{45}
\end{eqnarray}
Thereby, the decay time $T\sim1/\gamma$ of the damping field may be estimated by simultaneously estimating $k^2\sim1/L^2$. As a result, we have
\begin{eqnarray}
\left(\frac{T}{\mu\sigma L^2}\right)^3-\left(1+\frac{\tau}{\mu\sigma L^2}\right)\left(\frac{T}{\mu\sigma L^2}\right)^2+\frac{\epsilon/\sigma}{\mu\sigma L^2}\frac{T}{\mu\sigma L^2}-\frac{\epsilon/\sigma}{\mu\sigma L^2}\frac{\tau}{\mu\sigma L^2}=0.
\label{46}
\end{eqnarray}
To simplify the notation, first we name the characteristic time scale of conduction as
\begin{eqnarray}
\tau_\ast=\frac{\epsilon}{\sigma},
\label{47}
\end{eqnarray}
and, subsequently, normalize all times to the diffusion time,
\begin{eqnarray}
\hat{\tau}=\frac{\tau}{\mu\sigma L^2},\;\;\;\;\;\;\;\;\;\;\hat{\tau}_\ast=\frac{\tau_\ast}{\mu\sigma L^2},\;\;\;\;\;\;\;\;\;\;\hat{T}=\frac{T}{\mu\sigma L^2}.
\label{48}
\end{eqnarray}
Therefore, we obtain
\begin{eqnarray}
\hat{T}^3-\left(1+\hat{\tau}\right)\hat{T}^2+\hat{\tau}_\ast\hat{T}-\hat{\tau}_\ast\hat{\tau}=0.
\label{49}
\end{eqnarray}
Eq. (\ref{49}) is a cubic algebraic equation in $\hat{T}$. However, by considering the transformation \cite{Birkhoff}
\begin{eqnarray}
\hat{T}=\hat{T}_\ast+\left(\frac{1+\hat{\tau}}{3}\right)+\frac{1}{\hat{T}_\ast}\left[\left(\frac{1+\hat{\tau}}{3}\right)^2-\frac{\hat{\tau}_\ast}{3}\right],
\label{50}
\end{eqnarray}
one may easily check that $\hat{T}_\ast^3$ satisfies the quadratic algebraic equation
\begin{eqnarray}
\left(\hat{T}_\ast^3\right)^2-2\left[\left(\frac{1+\hat{\tau}}{3}\right)^3-\frac{\hat{\tau}_\ast}{3}\left(\frac{1}{2}-\hat{\tau}\right)\right]\hat{T}_\ast^3+\left[\left(\frac{1+\hat{\tau}}{3}\right)^2-\frac{\hat{\tau}_\ast}{3}\right]^3=0,
\label{51}
\end{eqnarray}
whose general solutions are given by
\begin{eqnarray}
&&\hat{T}_\ast^3=\left(\frac{1+\hat{\tau}}{3}\right)^3-\frac{\hat{\tau}_\ast}{3}\left(\frac{1}{2}-\hat{\tau}\right)\pm\sqrt{\frac{\hat{\tau}_\ast}{3}}
\nonumber\\&&\left\{\frac{\hat{\tau}_\ast}{3}+\frac{1}{3}\left[\left(\frac{5-3\sqrt{3}}{4}-\hat{\tau}\right)\left(\frac{5+3\sqrt{3}}{4}-\hat{\tau}\right)+\left(\frac{1}{4}-2\hat{\tau}\right)^{3/2}\right]\right\}^{1/2}\nonumber\\&&\left\{\frac{\hat{\tau}_\ast}{3}+\frac{1}{3}\left[\left(\frac{5-3\sqrt{3}}{4}-\hat{\tau}\right)\left(\frac{5+3\sqrt{3}}{4}-\hat{\tau}\right)-\left(\frac{1}{4}-2\hat{\tau}\right)^{3/2}\right]\right\}^{1/2}.
\label{52}
\end{eqnarray}
We see that there are three possible pairs of values for $\hat{T}_\ast$. Each pair is given by the plus or minus signs in Eqs. (\ref{52}). However, by substituting a chosen pair in Eq. (\ref{50}), one may easily check that the same value for $\hat{T}$ is obtained \cite{Birkhoff}. Thereby, as expected, all cubic algebraic equations, Eqs. (\ref{49}), (\ref{46}) and (\ref{45}), have three solutions. We pass now to the analyzis of important particular situations.

\subsection{Neglect of inertia and displacement current}
By neglecting both inertial effects and displacement current term, one expects the classical result for the decay time of the magnetic field to be recovered. Indeed, by taking the limits $\hat{\tau}\rightarrow0$ and $\hat{\tau}_\ast\rightarrow0$, simultaneously, in Eqs. (\ref{52}), we get $\hat{T}_\ast=1/3$, so that Eq. (\ref{50}) gives $\hat{T}=1$. In other words, the decay time of the damping field is given by the diffusion time, $T=\mu\sigma L^2$ \cite{Landau}.

\subsection{Influence of displacement current}
We now relax the approximation $\hat{\tau}_\ast\rightarrow0$ although still keep the limit $\hat{\tau}\rightarrow0$ in Eqs. (\ref{52}) to get
\begin{eqnarray}
\hat{T}_\ast=\left\{\left(\frac{1}{3}\right)^3-\frac{\hat{\tau}_\ast}{3}\left(\frac{1}{2}\right)\pm\frac{\hat{\tau}_\ast}{3}\left[\frac{\hat{\tau}_\ast}{3}-\frac{1}{3}\left(\frac{1}{4}\right)\right]^{1/2}\right\}^{1/3},
\label{53}
\end{eqnarray}
so that Eq. (\ref{50}) gives
\begin{eqnarray}
\hat{T}=\hat{T}_\ast+\frac{1}{3}+\frac{1}{\hat{T}_\ast}\left[\left(\frac{1}{3}\right)^2-\frac{\hat{\tau}_\ast}{3}\right].
\label{54}
\end{eqnarray}
Two limiting cases naturally arise.

\subsubsection{Small displacement current}
First, let us suppose that the influence of displacement current term is small in comparison with magnetic diffusion effects. In other words, let us assume that the approximation $\hat{\tau}_\ast\ll1$ holds in Eqs. (\ref{53}) to get
\begin{eqnarray}
\hat{T}_\ast=\frac{1}{3}-\frac{\hat{\tau}_\ast}{2}\pm\imath\frac{\hat{\tau}_\ast}{2}\left(\frac{1}{\sqrt{3}}\right),
\label{55}
\end{eqnarray}
so that Eq. (\ref{54}) gives
\begin{eqnarray}
\hat{T}=1-\hat{\tau}_\ast.
\label{56}
\end{eqnarray}
We see that the diffusion time is slightly decreased by the conduction time as a consequence of assuming a small influence of displacement current,
\begin{eqnarray}
T=\mu\sigma L^2-\frac{\epsilon}{\sigma}.
\label{57}
\end{eqnarray}
For all practical purposes, the conduction time is normally neglected in Eq. (\ref{57}), and the decay time of the damping field is assumed again to be given simply by the diffusion time.

\subsubsection{Large displacement current}
Second, let us suppose that the influence of displacement current is large in comparison with magnetic diffusion effects. In other words, let us assume that the approximation $\hat{\tau}_\ast\gg1$ holds in Eqs. (\ref{53}) to get
\begin{eqnarray}
\hat{T}_\ast=\pm\sqrt{\frac{\hat{\tau}_\ast}{3}},
\label{58}
\end{eqnarray}
so that Eq. (\ref{54}) gives
\begin{eqnarray}
\hat{T}=\frac{1}{3}.
\label{59}
\end{eqnarray}
We see that the diffusion time is strongly suppressed to a third of its classical value as a consequence of assuming a large influence of displacement current,
\begin{eqnarray}
T=\frac{\mu\sigma L^2}{3}.
\label{60}
\end{eqnarray}
Note that, although the influence of displacement current is arguably very much larger than magnetic diffusion effects, the decay time does not depend on the conduction time. This interesting situation could arise in the limit of typically very small samples. However, it should be difficult to be observed, under normal conditions, given the smallness of the length scale.

\subsection{Influence of inertia}
Finally, we relax the approximation $\hat{\tau}\rightarrow0$ although still keep the limit $\hat{\tau}_\ast\rightarrow0$ in Eqs. (\ref{52}) to get
\begin{eqnarray}
\hat{T}_\ast=\frac{1+\hat{\tau}}{3},
\label{61}
\end{eqnarray}
so that Eq. (\ref{50}) gives
\begin{eqnarray}
\hat{T}=1+\hat{\tau}.
\label{62}
\end{eqnarray}
We see that the diffusion time is increased by the relaxation time as a consequence of inertial effects,
\begin{eqnarray}
T=\mu\sigma L^2+\tau.
\label{63}
\end{eqnarray}
Although this is the same result of \cite{Cuevas}, we stress here that, in order to derive it, the subsidiary condition $\gamma\tau\ll1$ does not need at all to be imposed, contrarily as claimed by those authors. Moreover, we note that the effects of displacement current term on decay time, Eqs. (\ref{57}) and (\ref{60}), are opposite to that of inertial effects, Eq. (\ref{63}). Such a result could explain why, under ordinary circumstances, the decay time of damping fields scales approximately as the diffusion time. As a matter of fact, Eq. (\ref{63}) does not rule out the possibility of considering the particular situation for which magnetic diffusion effects are neglibible in comparison with inertial ones, $\mu\sigma L^2\ll\tau$. In that case, the decay time would scale approximately as the relaxation time,
\begin{eqnarray}
T=\tau.
\label{64}
\end{eqnarray}
That interesting situation could arise in the limit of typically very small samples (see also \cite{Cuevas}): at the nano scale, for instance, provided quantum mechanical effects were unimportant.

\section{conclusion}
In this work, new results for space attenuation of electromagnetic waves and time damping of magnetic fields in rigid conducting media have been derived by taking into account the combined effects of inertia, due to charge carriers, and displacement current term. The influence of inertia has been realized by the introduction of the relaxation time for the current density, through a suitable generalization of Ohm's law. The main conclusions of this article can be summarized as follows:

\vspace{.5cm}

\noindent(1) The classical notions of poor and good conductors have been extended in the framework of an effective electric conductivity. That quantity depends on both wave frequency and relaxation time [see Eq. (\ref{9})].

\vspace{.5cm}

\noindent(2) It has been found that the space attenuation for good conductors at high frequency regimes depends on the relaxation time only [see the second of Eqs. (\ref{40})]. In other words, the penetration depth must saturate to a minimum value at sufficiently high frequencies. Such a result leads to the possibility of measurement of the relaxation time (see Appendix B).

\vspace{.5cm}

\noindent(3) The overall problem of determining the decay time of a magnetic field which is suddenly removed from a rigid conductor has been solved on very general grounds, that is, by taking into account the conjugate influences of magnetic diffusion, displacement current and inertia due to charge carriers. The solutions are given by the roots of a cubic algebraic equation involving all the relevant parameters [see Eq. (\ref{46})].

\vspace{.5cm}

\noindent(4) We have shown that the actions of displacement current term and inertia due to charge carriers on damping of magnetic fields are opposite to each other [compare Eqs. (\ref{57}) and (\ref{60}) with (\ref{63})]. In particular, this result explains why, under normal conditions, the classical decay time of damping fields scales approximately as the diffusion time.

\vspace{.5cm}

\noindent(5) At very small length scales (at nano scales, for instance), provided effects due to quantum mechanics were unimportant, it is possible that the decay time could be given either by a third of the diffusion time [see Eq. (\ref{60})] or by the relaxation time [see Eq. (\ref{64})], depending on whether the displacement current term or inertia due to charge carriers, respectively, would prevail on magnetic diffusion. However, under ordinary circumstances, the former situation should not be easily observable given the smallness of the length scale.

\acknowledgments{The authors are grateful to Rose C. Santos and M. P. M. Assun\c{c}\~ao for helpful discussions. JASL is partially supported by CNPq and FAPESP (Brazilian Research Agencies) under grants  304792/2003-9 and 04/13668-0, respectively.}

\appendix

\section{Ohm's law and Drude's model}
Consider the action of a microscopic electric field $\vec{\mathcal{E}}$ on a conduction electron of negative charge $e$ and mass $m$ in a typical metal. The resulting velocity $\vec{v}$ of the electron is supposed to satisfy the stochastic Langevin's equation \cite{Risken,Gardiner,Kampen},
\begin{eqnarray}
m\frac{d\vec{v}}{dt}=e\vec{\mathcal{E}}-\nu\vec{v}+\vec{\eta},
\label{a1}
\end{eqnarray}
where $\vec{\eta}$ is a random force, commonly referred to as the noise force, and $\nu\vec{v}$ is a viscous force, with $\nu$ denoting a viscosity coefficient. By regarding collisions of electrons with ions as the only possible dissipative process (classical collisional regime), the viscosity coefficient of the medium is assumed to be given by \cite{Risken,Gardiner,Kampen}
\begin{eqnarray}
\nu=\frac{m}{\tau_{\rm c}},
\label{a2}
\end{eqnarray}
where $\tau_{\rm c}$ is know as the classical collision time. The resulting current density flowing in the conductor is defined as $\vec{J}=Ne<\vec{v}>$, where $N$ is the number density of electrons and the angular brackets denote an ensemble average. Now, the ensemble averaged noise force vanishes \cite{Risken,Gardiner,Kampen}, $<\vec{\eta}>=0$, and the macroscopic electric field $\vec{E}$ is given by the ensemble average of the microscopic electric field, $\vec{E}=<\vec{\mathcal{E}}>$. Therefore, Langevin's equation, Eq. (\ref{a1}), leads to
\begin{eqnarray}
\left(1+\tau_{\rm c}\frac{d}{dt}\right)\vec{J}=\sigma_{\rm D}\vec{E},
\label{a3}
\end{eqnarray}
where we identify the classical Drude's electric conductivity \cite{Born,Klein}
\begin{eqnarray}
\sigma_{\rm D}=\frac{Ne^2\tau_{\rm c}}{m}.
\label{a4}
\end{eqnarray}
We see that Eq. (\ref{a3}) corresponds to the generalized Ohm's law of Sec. I, provided  $\tau\sim \tau_{\rm c}$ and $\sigma\sim \sigma_{\rm D}$ in Eq. (\ref{2}).

\section{Lower limit for relaxation time}
For our purposes, we rewrite the dispersion relation Eq. (\ref{12}) in a more convenient manner,
\begin{eqnarray}
k^2=\omega^2\mu\epsilon\epsilon_{\rm r},
\label{b1}
\end{eqnarray}
where we identify the complex relative electric permittivity
\begin{eqnarray}
\epsilon_{\rm r}=1+\imath\frac{1}{\epsilon\omega}\frac{\sigma}{\sqrt{1+\omega^2\tau^2}}\exp{\left(\imath\varphi\right)},
\label{b2}
\end{eqnarray}
with the time dephasing angle $\varphi$ given by Eq. (\ref{10}), and recognizing the effective electric conductivity $\sigma_{\rm eff}$ [see Eq. (\ref{9})] as the second fraction on the right hand side of the above equation. In the limit $\tau\rightarrow0$, Eq. (\ref{b2}) recovers the expression for the classical relative permittivity \cite{Born,Klein}. However, in the high frequency regime $\omega\tau\gg1$, $\varphi\cong\pi/2$ [see Eq. (\ref{32})] and the relative permittivity becomes real. Moreover, in the good conductor limit $\sigma_{\rm eff}\gg\epsilon\omega$ (see Sec. III), $\epsilon_{\rm r}$ becomes negative and $k$, purely imaginary [see the second of Eqs. (\ref{40})]. By binomial expanding the expression $\sqrt{1+\omega^2\tau^2}$ to the term of $\mathcal{O}\left(\omega^2\tau^2\right)\gg1$, Eq. (\ref{b2}) approaches
\begin{eqnarray}
\epsilon_{\rm r}=1-\frac{1}{\omega^2}\frac{\sigma}{\epsilon\tau}+\frac{1}{\omega^4}\frac{\sigma}{2\epsilon\tau^3}.
\label{b3}
\end{eqnarray}
Again, since $\epsilon_{\rm r}<0$, $\omega$ must satisfy the condition
\begin{eqnarray}
\omega^4-\frac{\sigma}{\epsilon\tau}\omega^2+\frac{\sigma}{2\epsilon\tau^3}<0.
\label{b4}
\end{eqnarray}
In other words, $\omega$ must lie in the region
\begin{eqnarray}
\sqrt{\frac{\sigma}{2\epsilon\tau}}\left[1-\sqrt{1-\frac{2\epsilon}{\sigma\tau}}\right]^{1/2}<\omega<\sqrt{\frac{\sigma}{2\epsilon\tau}}\left[1+\sqrt{1-\frac{2\epsilon}{\sigma\tau}}\right]^{1/2}.
\label{b5}
\end{eqnarray}
Also, since $\omega$ is real, $\tau$ must satisfy the condition
\begin{eqnarray}
\tau>\frac{2\epsilon}{\sigma}.
\label{b6}
\end{eqnarray}
Again, by assuming that the relaxation and collision times have the same order of magnitude (see Appendix A), and introducing the plasma frequency \cite{Born,Klein}
\begin{eqnarray}
\omega_{\rm p}=\sqrt{\frac{Ne^2}{m\epsilon}},
\label{b7}
\end{eqnarray}
$\tau_{\rm c}$ must satisfy the condition
\begin{eqnarray}
\tau_{\rm c}>\frac{\sqrt{2}}{\omega_{\rm p}}.
\label{b8}
\end{eqnarray}
For a typical good metal like copper, $N\cong8.4\times10^{22}\;{\rm cm}^{-3}$ so that $\omega_{\rm p}\cong1.6\times10^{16}\;{\rm s}^{-1}$. Since $\tau_{\rm c}\cong2.7\times10^{-14}\;{\rm s}$, condition (\ref{b8}) is satisfied. Direct observations of plasmons, like those in connection with thin metallic films \cite{Powell}, are always hard to be accomplished. However, the possibility of their indirect detection during various electronic processes should not be underestimated \cite{Dragila,Tsuei,Tediosi}.

\end{document}